\documentclass[12pt]{article}
\usepackage[pdftex]{graphicx}
\usepackage[pdftex]{color}
\usepackage{times,amsmath,dcolumn,wrapfig}
\renewcommand{\d}{\,\text{d}}
\newcommand{\beq}{\begin{equation}}
\newcommand{\eeq}{\end{equation}}

\title{TRI-BN-25-9: Cyclotron Mirror Inflector}
\author{Rick Baartman, TRIUMF}
\date{April, 2025}
\begin{document}
\maketitle
\raggedright
\begin{abstract}
The device for injecting into a compact cyclotron is known as an inflector. The oldest simplest version is an electrostatic mirror. From the equations of motion through I derive the geometry of the mirror. Also the transfer matrix is given in symplectic form.\end{abstract}

\section{Introduction}
An electrostatic mirror is a localized electric field such as is often used to steer a particle beam. Usually this is simply two parallel plates, one of which has at least one gridded aperture to allow the beam to enter. For injecting into a cyclotron, these are oriented at an angle with respect to the vertical central axis of the cyclotron. Without a magnetic field, this angle would be $\pi/4$ to achieve a $\pi/2$ bend. However with the magnetic field, the beam will begin to rotate about the vertical axis and attain some motion in the direction orthogonal to the electric and magnetic fields as soon as it has gained some angle with respect to vertical. To bring the beam onto the median plane with no remaining momentum component in the vertical direction, the mirror must be at an angle steeper than $45^\circ$.
\begin{figure}[htbp]
\begin{center}
\includegraphics[width=.5\textwidth]{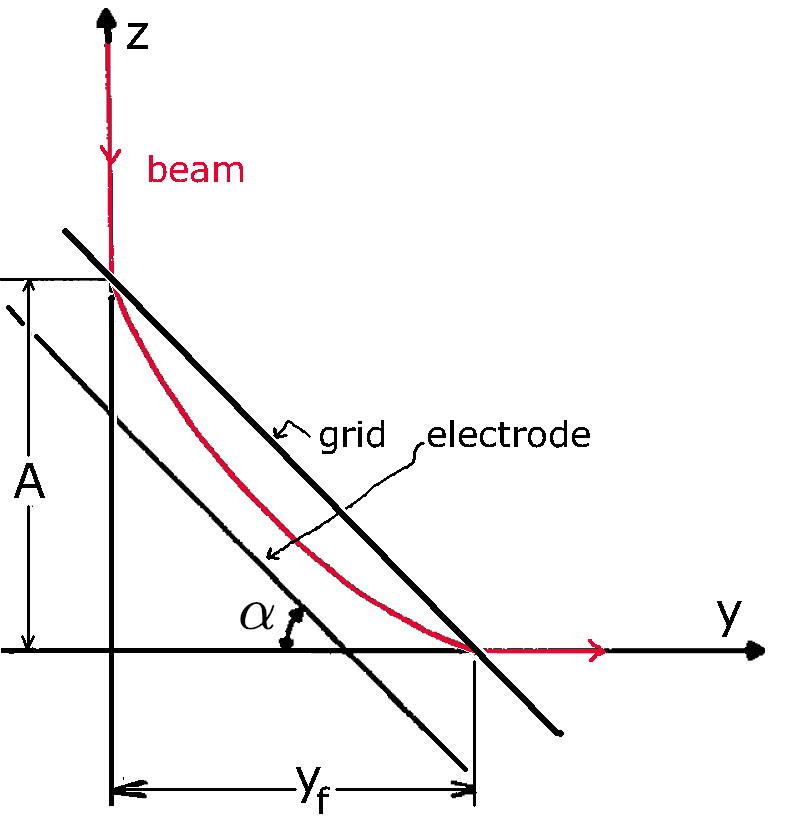}
\caption{Side view of mirror inflector. Electric field is in the $y$-$z$-plane. Not visible in this projection is that due to the magnetic field $B$ in the $z$ direction, the beam also moves along $x$, toward or away from the viewer depending upon the sign of $B$.}
\label{fig:mirror}
\end{center}
\end{figure}
The following analysis relies on the previous work of Bellomo et al.\cite{bellomo1983feasibility}, but with somewhat simpler notation.

As the electric field is not continually orthogonal to the beam (as it is for example in a spiral inflector), the beam's kinetic energy varies through its trajectory. It is therefore simpler and more convenient to base the equations of motion on time as independent variable. 

\section{Equations of Motion}
Beam arriving at $z=A$, $A$ being inflector height, see Fig.\,\ref{fig:mirror}, is to be inflected so that when it reaches the exit of the inflector, it is on the median plane $z=0$ and its vertical momentum is zero. Since the vertical magnetic field exerts no vertical force on the beam, only the vertical component of the electric force has any effect on the motion in the $z$ direction: 
\beq 
m\ddot{z}=q{\cal E}\cos\alpha\mbox{, or, }\ddot{z}=a_z.
\eeq 
(dots denote time derivative). Acceleration $a_z=q{\cal E}\cos\alpha/m$ is a constant. This is easily integrated using initial condition $z(0)=A$, $\dot{z}(0)=-v_0$: 
\beq 
z=\frac{1}{2}a_zt^2-v_0t+A.
\eeq 
Let $t_{\rm f}$ be the time at exit, then we have $v_0=a_zt_{\rm f}$ and $A=\frac{1}{2}v_0t_{\rm f}$, so 
\beq 
z=A\left(1-\frac{t}{t_{\rm f}}\right)^2.
\eeq

We can now express the inflector height $A$ in terms of the electric field:
\beq
A=\frac{1}{2}a_zt_{\rm f}^2=\frac{v_0^2/2}{a_z}=\frac{mv_0^2/2}{q{\cal E}\cos\alpha}=\frac{V_0}{{\cal E}\cos\alpha},
\eeq
where $V_0$ is the particles' energy per charge, usually the potential of the ion source terminal with respect to ground.

The Lorentz force equation gives us the $x$ equation: \beq m\ddot{x}=-qB\dot{y}\mbox{, or, }\ddot{x}=-\omega\dot{y},\eeq where the frequency $\omega$ is simply the cyclotron frequency \beq \omega=\frac{v_0}{\rho}=\frac{qB}{m}.\eeq Similarly, \beq \ddot{y}=a_z\tan\alpha+\omega\dot{x}.\eeq

It's handy to define a dimensionless time $\tau=\omega t$. In other words, $\tau$ is the angle of the circle that a particle on the median plane would travel. Note that scaled time at exit from mirror is \beq\tau_{\rm f}=\omega t_{\rm f}=2A/\rho:=2k,\eeq
introducing an inflector parameter $k=A/\rho$ similar to the one used for the spiral inflector.

With dots now being derivatives with respect to $\tau$, we have 
\begin{eqnarray}
\ddot{x}&=&-\dot{y}\nonumber\\
\ddot{y}&=&\dot{x}+\frac{\rho}{2k}\tan\alpha\label{eomn}
\end{eqnarray}

For initial conditions $x=y=\dot{x}=\dot{y}=0$, the solution is 
\begin{eqnarray}
x&=&-\frac{\rho}{2k}\tan\alpha\,(\tau-\sin\tau)\nonumber\\
y&=&\frac{\rho}{2k}\tan\alpha\,(1-\cos\tau).\label{onaxis}
\end{eqnarray}
These are in fact the parametric equations of a cycloid. (Not a surprise: it is well known that particles in crossed electric and magnetic fields travel on a cycloid in the direction of $\vec{\cal E}\times\vec{B}$). On these terms, $z$ is:
\begin{eqnarray}
z&=&A\left(1-\frac{\tau}{2k}\right)^2.
\end{eqnarray}

We now derive the condition on the angle $\alpha$. Since $y(2k)=y_{\rm f}$ must satisfy $A/y_{\rm f}=\tan\alpha$ (see Fig.\,\ref{fig:mirror}), we find:
\beq
A=\frac{\rho}{2k}\tan^2\alpha\,(1-\cos(2k)),
\eeq
or (believe it or not),
\beq
\tan\alpha=\frac{k}{\sin k}.\label{believe}
\eeq
If $A$ is small on the scale of $\rho$, $\alpha$ is near $45^\circ$. At the extreme, if $k=\pi/2$, $y$ attains an extremum value and the beam can exit out the side of the plates, $\alpha$ is $57.5^\circ$. But this is unworkable as we shall see, since it places the beam far off centre. Generally, $0<|k|<1$ and so $45^\circ<\alpha<50^\circ$.

\section{Centring}
Given these equations of motion, we can plot the trajectory. The beam is not well-centred in the cyclotron, if it enters the inflector on the cyclotron's axis. It is thus often necessary to place the inflector off axis. We can find the centring error as follows.
\begin{figure}[htbp]
\begin{center}
\includegraphics[width=.8\textwidth]{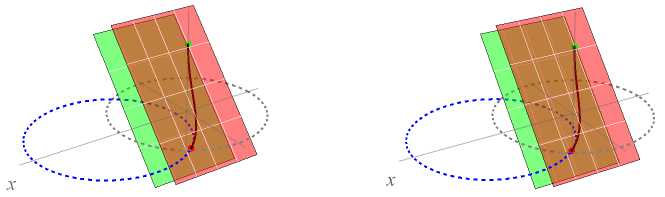}
\caption[Caption for LOF]{The beam's trajectory for the case $k=0.9$. Two perspectives are shown as a stereoscopic view.\footnotemark\ The mirror plates are shown: upper (red) electrode is shown as semi-transparent to make visible the trajectory, but does in fact have to be gridded to allow the beam into and out of the mirror. The dashed circles are in the $x$-$y$-plane; the grey one is the centred orbit and the blue one is the actual orbit if uncorrected.}
\label{fig:mirrororbits}
\end{center}
\end{figure}
\begin{figure}[htbp]
\begin{center}
\includegraphics[width=\textwidth]{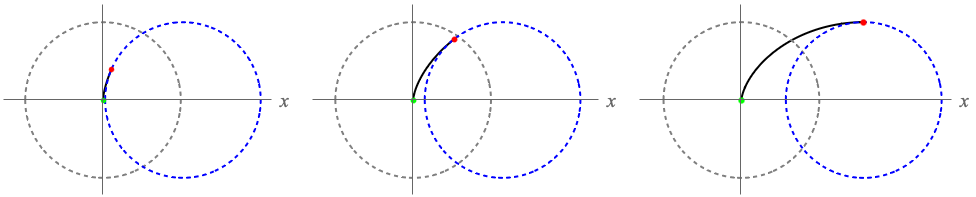}
\caption{Three cases in overhead view ($x$-$y$ projection). Resp.\ $k=.1,.9,\pi/2$; the centre case is same as in Fig.\,\ref{fig:mirrororbits}.}
\label{fig:mirrorOH}
\end{center}
\end{figure}
\footnotetext{See Appendix A on how to view in 3D using such stereoscopic pairs.}

From the equations of motion, we find
\begin{eqnarray}
x_{\rm f}&=&\rho(-\cos k+k\csc k)=\rho(-\cos k+\tan\alpha)\\
y_{\rm f}&=&\rho\sin k=\frac{A}{\tan\alpha}.
\end{eqnarray}
Since the orbits in the uniform field are circles\footnote{left as an exercise ...}, we have:
\begin{eqnarray}
x_{\rm c}&=&x_{\rm f}+\dot{y}_{\rm f}=\rho\tan\alpha\approx\rho\nonumber\\
y_{\rm c}&=&y_{\rm f}-\dot{x}_{\rm f}=0;\label{exittocentre}
\end{eqnarray}
The orbit is displaced `sideways' with respect to the electric field, approximately one radius $\rho$ away from the entry point of the inflector. This is shown pictorially in the plots of the $x$-$y$ projection in Fig.\,\ref{fig:mirrorOH}. It means that in order to centre the beam in the cyclotron, the beam and inflector needs to be displaced in the direction orthogonal to the electric field by about one $\rho$. In the Bellomo et al.\ design\cite{bellomo1983feasibility}, this issue is discussed at some length. Two deflectors are used to place the beam 10\,mm off axis, for $\rho=8.3$\,mm, $\rho\tan\alpha=9.1$\,mm.

\section{Optics}
The equations of motion \ref{eomn} can be solved for the general case of off-axis and off-momentum initial conditions. Bellomo et al.\cite{bellomo1983feasibility} have already done this and we will not repeat their work here. A transformation needs to be made because the Cartesian coordinates that we started with are not the usual Frenet-Serret ones; they are at the start, but not at the exit of the inflector. There, the $x$ and $y$ need to be transformed to deviations with respect to the central trajectory. As usual, we define $x$ as the radially outward increment, and $z$ as the vertical. But to avoid confusion with the Cartesian ones, we label them with the subscript `FS'.

The Bellomo matrix that takes beam from the beamline (`BL') to the cyclotron (`cy') becomes, in our notation,
\small\beq
\begin{pmatrix}
x_{\rm FS}/\rho \\x'_{\rm FS}\\z_{\rm FS}/\rho\\z'_{\rm FS}\\\zeta/\rho\\\Delta p/p\\
\end{pmatrix}_{\rm cy}  =
\begin{pmatrix}
   \cos k &                  2\sin k & \sin k & 0 & 0 & 0 \\
{-\sin k} &   \frac{\cos 2k}{\cos k} & \cos k & 0 & 0 & \tan k\\
        0 &                        0 & -\frac{k}{\sin k} & 0 & 0 & -2k\\
        0 & -\frac{\sin^2k}{k\cos k} & 0 & -\frac{\sin k}{k} & 0 & \frac{\tan k}{k}-1 \\
  -\sin k &                        0 & \cos k-\frac{k}{\sin k} & 2\sin k & 1 & 0 \\
        0 &                        0 &                       0 & 0 & 0 & 1 \\
                                       \end{pmatrix} 
\begin{pmatrix}
x/\rho \\x' \\y/\rho \\ y' \\ \zeta/\rho \\ \Delta p/p \\
\end{pmatrix}_{\rm BL}\label{nonsymp}
\eeq\normalsize
Note that to simplify the matrix, we normalize lengths by $\rho$, as done by Bellomo et al. Nevertheless, there are differences with their matrix that need clarification. 
\begin{itemize}
\item We interchange derivatives with respect to $\tau$ and with respect to $s$, the distance along the reference orbit, as $\tau=\omega t=vt/\rho=s/\rho$, $\dot{x}\equiv\frac{\d x}{\d\tau}=\rho\frac{\d x}{\d s}$, and so $x'=\dot{x}/\rho$, and similarly for $y$ and for the FS coordinates. 
\item The longitudinal coordinate $\Delta\tau$ is opposite in sign to the canonical one, which we denote $\zeta$ here; $\zeta= -v_0\Delta t=-\omega\rho\Delta t$, or $\zeta/\rho=-\Delta\tau$. The result is that the matrix here has changed sign for row 5 and column 5, compared with the Belomo et al.\ matrix. Mathematically, this is because in the $s$ as independent variable case, the longitudinal coordinate is not $t$, but $-t$. Intuitively, it is because a particle ahead of the reference particle ($\zeta>0$), has earlier arrival time ($\Delta t<0$).
\item All incidences of $\tan\alpha$ have been replaced with $\frac{k}{\sin k}$ as in eq.\,\ref{believe}.
\end{itemize}

If we apply the usual tests for symplecticity, it fails. In particular, columns 1 and 3 give determinant 1 instead of zero\footnote{see Appendix B for explanation and a brief `do in your head' technique of this test.}. This is because the starting coordinates $(x',y')$ are not canonical momenta ($\vec{P}=\vec{p}+q\vec{A}$). Entry into the cyclotron's magnetic field on axis will achieve this same transformation, which is effectively an entry into a solenoidal field. The vector potential is 
\beq
(A_x,A_y)=\frac{B}{2}(-y,x).
\eeq
At the inflector exit, on the other hand, the vector potential is zero. The transformation is therefore
\small\beq
\begin{pmatrix}
x/\rho \\x' \\y/\rho \\ y' \\ \zeta/\rho \\ \Delta p/p \\
\end{pmatrix}_{\rm BL}
  =
\begin{pmatrix}
           1 & 0 &            0 & 0 & 0 & 0 \\
           0 & 1 & -\frac{1}{2} & 0 & 0 & 0 \\
           0 & 0 &            1 & 0 & 0 & 0 \\
 \frac{1}{2} & 0 &            0 & 1 & 0 & 0 \\
           0 & 0 &            0 & 0 & 1 & 0 \\
           0 & 0 &            0 & 0 & 0 & 1 \\
\end{pmatrix} 
\begin{pmatrix}
x/\rho \\P_x/p \\y/\rho \\ P_y/p \\ \zeta/\rho \\ \Delta p/p \\
\end{pmatrix}_{\rm BL}\label{cantran}
\eeq\normalsize

Combining eqs.\,\ref{nonsymp},\ref{cantran},
\small\beq
\begin{pmatrix}
x_{\rm FS}/\rho \\x'_{\rm FS}\\z_{\rm FS}/\rho\\z'_{\rm FS}\\\zeta/\rho\\\Delta p/p\\
\end{pmatrix}_{\rm cy}  =
\begin{pmatrix}
   \cos k &                  2\sin k & 0 & 0 & 0 & 0 \\
-\sin k &   \frac{\cos 2k}{\cos k} & \frac{1}{2\cos k} & 0 & 0 & \tan k\\
        0 &                        0 & -\frac{k}{\sin k} & 0 & 0 & -2k\\
-\frac{\sin k}{2k} & -\frac{\sin^2k}{k\cos k} & \frac{\sin^2k}{2k\cos k} & -\frac{\sin k}{k} & 0 & \frac{\tan k}{k}-1 \\
        0 &                        0 & \cos k-\frac{k}{\sin k} & 2\sin k & 1 & 0 \\
        0 &                        0 &                       0 & 0 & 0 & 1 \\
                                       \end{pmatrix} 
\begin{pmatrix}
x/\rho \\P_x/p \\y/\rho \\ P_y/p \\ \zeta/\rho \\ \Delta p/p \\
\end{pmatrix}_{\rm BL}\label{mirsymp}
\eeq\normalsize
The student can verify that this matrix is indeed symplectic.

\bibliographystyle{elsarticle-num}
\bibliography{/Users/baartman/AllDN/Baartman,/Users/baartman/AllDN/AllDN,/Users/baartman/AllDN/Others}

\section*{Appendix A: 3D viewing using stereoscopic pairs}
The pairs of figures are to enable you, the reader to view a 3-dimensional image. This is done by having two views from slightly different angles as would occur when your two eyes actually see a real 3D object. Note that the figures here are in a `parallel-view', not a `cross-view', which would requiring crossing one's eyes. 

To view them, imagine the object is far in the distance so that you see 4 images (2 from each eye), and convince the eyes to superimpose the centre two. Once that is achieved, try to focus without re-converging the eyes. You may need to adjust the magnification on screen so that the two images are no farther apart than your PD (pupillary distance of the eyes).

An alternative and more foolproof method is to place a barricade such as a sheet of paper between the eyes and extending to the screen or page, between the images; the intent is that the left eye see only the left image and the right eye see only the right image.

\section*{Appendix B: How to do symplectic test for $6\times6$ matrix.}
First, envision the 6 columns. There will be 15 tests: the first involves columns 1,2, then columns 1,3, then 1,4; 1,5; 1,6. That's 5 tests. Next do the same for 2,3; 2,4; 2,5; 2,6. Then 3,4; 3,5; 3,6. Then 4,5; 4,6. Lastly, 5,6.

Each test is as follows: Divide the two columns into three $2\times2$ matrices: rows 1,2, rows 3,4, rows 5,6. Find the three determinants, add them together. This sum should be exactly zero for 12 of the cases; only the column 1,2 case, the 3,4 case, the 5,6 case sum to exactly 1.

To be more mathematical, let us say we are dealing with columns $i$ and $j$. Then take the sum: $M_{1i}M_{2j}-M_{1j}M_{2i}+M_{3i}M_{4j}-M_{3j}M_{4i}+M_{5i}M_{6j}-M_{5j}M_{6i}$. This is 1 for $(i,j)=(1,2)$ or $(3,4)$ or $(5,6)$, and 0 for the other twelve.

\includegraphics[width=0.4\textwidth]{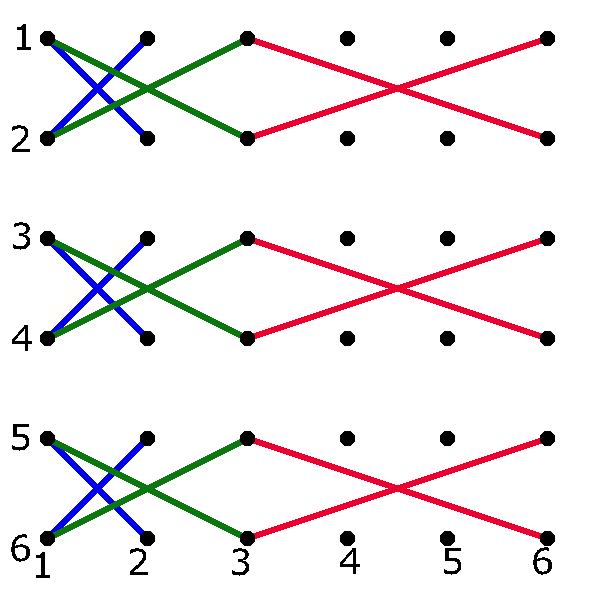}
This is pictorially shown in the figure. Black dots represent matrix elements, coloured lines connect the elements that are multiplied together, left to right diagonally downward are positive and upward are negative terms. Only three examples are shown (showing all 15 would be too cluttered). In this example, blue is $(i,j)=(1,2)$ (should add to 1), green is $(i,j)=(1,3)$ (should be zero), red is $(i,j)=(3,6)$ (should add to zero).

\end{document}